# Effect of Misfit and Threading Dislocations on Surface Energies of PbTe-PbSe Interfaces

Emir Bilgili[1*], Nicholas Taormina[1], Yang Li[2], Adrian Diaz[3], Simon R. Phillpot[4], Youping Chen[1]

[1] Department of Mechanical and Aerospace Engineering, University of Florida, Gainesville, FL 32611, USA

[2] Department of Physics and Astronomy, MSN 3F3, George Mason University, Fairfax, VA 22030, USA

[3] Northrop Grumman Corporation, Linthicum Heights, MD 21090, USA

[4] Department of Materials Science and Engineering, University of Florida, Gainesville FL 32611, USA

## Abstract:

This work quantifies the effect of misfit and threading dislocations on the surface energies of PbTe-PbSe interfaces, with the defect structures of the interfaces being obtained from atomistic and multiscale simulations of their manufacturing processes. Simulation results show that direct bonding produces semi-coherent interfaces with two-dimensional misfit dislocation networks, while heteroepitaxial processes produce complex three-dimensional dislocation structures with both misfit and threading dislocations. Surface energies at these interfaces were determined by computing the interaction energies across these interfaces. Compared with coherent interfaces, directly bonded interfaces exhibit up to ~23% lower surface energy, while the surface energies of epitaxially grown interfaces can be nearly 50% lower. The results demonstrate the significant effects of dislocations on interfacial energy.

* *Corresponding author: emir.bilgili@ufl.edu*

Surface energy is a fundamental physical property that quantifies the energy required to disrupt intermolecular bonds when a surface is created and is a core concept across many science and engineering fields. For instance, in fracture mechanics, the surface energy determines the fracture toughness of a brittle material [1,2], and the cleavage energy of a crystalline material is twice the surface energy at a cleavage plane [3,4]. In theories of epitaxial growth, the surface energies of the overlayer and substrate, as well as their interfacial energy, dictate the growth mode [5,6]. In surface and interface science, surface energy determines how materials behave at a surface or an interface, influencing phenomena such as wetting or adhesion [7].

Existing computational approaches to determine surface or interfacial energies mainly rely on slab-based methods. Slab-based methods compute surface energy by creating a finite-thickness slab of a material's crystal structure and introducing a vacuum normal to the facet of interest which breaks periodicity along that direction, creating a surface. The surface energy is then calculated by taking the difference between the total energy of the slab and that of an equivalent amount of bulk crystal material [8]. Limitations of slab-based methods have been widely discussed [9,10]. Significant efforts have also been made to improve the reliability, efficiency, and accuracy of slab-based methods for surface energy calculations of elemental and arbitrary single-crystals [11–13].

Despite substantial progress in computational methods for determining surface energies of single-crystal materials, predicting the surface energy at an interface separating two dissimilar materials remains a critical challenge. This is because the structures of heterointerfaces are complex and highly dependent on the manufacturing processes during which they form. The structural complexity of physical interfaces, especially defects such as dislocations, are significantly difficult for analytical models and nanoscale methods such as density functional theory (DFT) or molecular dynamics (MD) to predict. As a result, a quantitative understanding of the surface energy at an interface, i.e., the interfacial energy, remains limited. For lattice mismatched heterostructures, the formation of dislocations during manufacturing is unavoidable. Yet, up to date, there are no quantitative studies on the effect of dislocations on the surface energies at interfaces.

The objective of this work is to demonstrate and quantify the effect of misfit and threading dislocations on interfacial energy. To do so, we first obtain the structures of the PbTe/PbSe and PbSe/PbTe interfaces through atomistic and multiscale simulation of their manufacturing processes, including direct bonding and heteroepitaxial growth. We then directly compute the interaction energies across the interfaces to determine the surface energies at these interfaces. We select the PbTe-PbSe system for this study because there is an interatomic potential available [14,15] that can reproduce the structure, type, and density of dislocations at the PbTe/PbSe interfaces [15–18] observed in experimental studies [19–21].

The direct wave bonding process is a widely used wafer bonding technique in semiconductor manufacturing [22]. To simulate the direct bonding process, first, each single-crystal material was relaxed to its equilibrium structure; second, the two relaxed crystals were stacked to form a heterostructure. In this work, the minimum interface size required is 19 PbTe unit-cells for every 20 PbSe unit-cells, corresponding very closely to the 0 K lattice mismatch of 4.94%. Third, these heterostructures were simulated under significant pressure perpendicular to the interface (150 Bar) at 300 K to promote bonding across the interface. Finally, these heterostructures were annealed to high temperature cycling between NPT and NVT ensembles from 300 → 1000 → 300 → 100 → 10 → 0.7 K, maintaining zero pressure. Any residual kinetic energy was removed using viscous damping with a damping coefficient of 0.1 eV·ps $Å^{-2}$, followed by energy minimization. Equilibrium was reached as the average force per atom fell below $10^{-12}$ eV $Å^{-1}$. The direct bonding process is found to produce atomically sharp semi-coherent interfaces with 2D misfit dislocation networks, as shown in Fig. 3.

In (111) PbTe and PbSe systems, the dislocation structures formed during heteroepitaxial growth depend sensitively on the size of the substrate [17], the length scales of which are usually not accessible to nanoscale methods such as DFT and MD. Therefore, to reach device-relevant length scales, multiscale simulations are performed using the concurrent atomistic continuum (CAC) method [23,24] which enables accurate modeling of substrates with finite-elements while the epitaxial growth process, including the formation of defects, is modeled with atomic resolution. We emphasize that, like MD, the CAC method does not require additional assumptions or empirical rules about mechanisms or material parameters (e.g., dislocation nucleation criteria) beyond an interatomic potential. Both the atomic and finite-element regions are governed by Newton's second law, and the resulting structures emerge naturally as a direct consequence of Newton's second law during a simulation. The epitaxial growth of PbTe/PbSe and PbSe/PbTe heterostructures in this work was simulated using CAC following procedures reported in our previous work [17]. Agreement between MD and CAC simulations of the heteroepitaxy for nanoscale systems was demonstrated in prior work [17,25]. As schematized in Fig. 1, the substrate was modeled using a multiscale scheme with a ~4 nm atomically resolved surface region and increasingly coarse finite-elements below with equivalent sizes of (4 × 4 × 4), (8 × 8 × 8), and (16 × 16 × 16) unit cells. The bottom layer was fixed, periodic boundary conditions were applied in-plane, and a vacuum region defined the free surface along the Z direction. In the atomically resolved region of the substrate, a 10–15 Å thick Langevin thermostat controlled the temperature, while the remaining specimen is governed by Newtonian mechanics. Deposition was simulated by randomly injecting epilayer atoms above the surface with Gaussian velocities corresponding to the growth temperature and a flux of ~0.1 monolayers $ns^{-1}$. A reflect particle boundary redirected floating atoms. Growth temperatures were 600 K for the PbTe-PbSe (100) and 900 K for PbTe-PbSe (111) systems. The heteroepitaxial growth process is found to produce interfaces with complex 3D dislocation structures that contain both misfit and threading dislocations, as shown in Fig. 2 and Fig. 3.

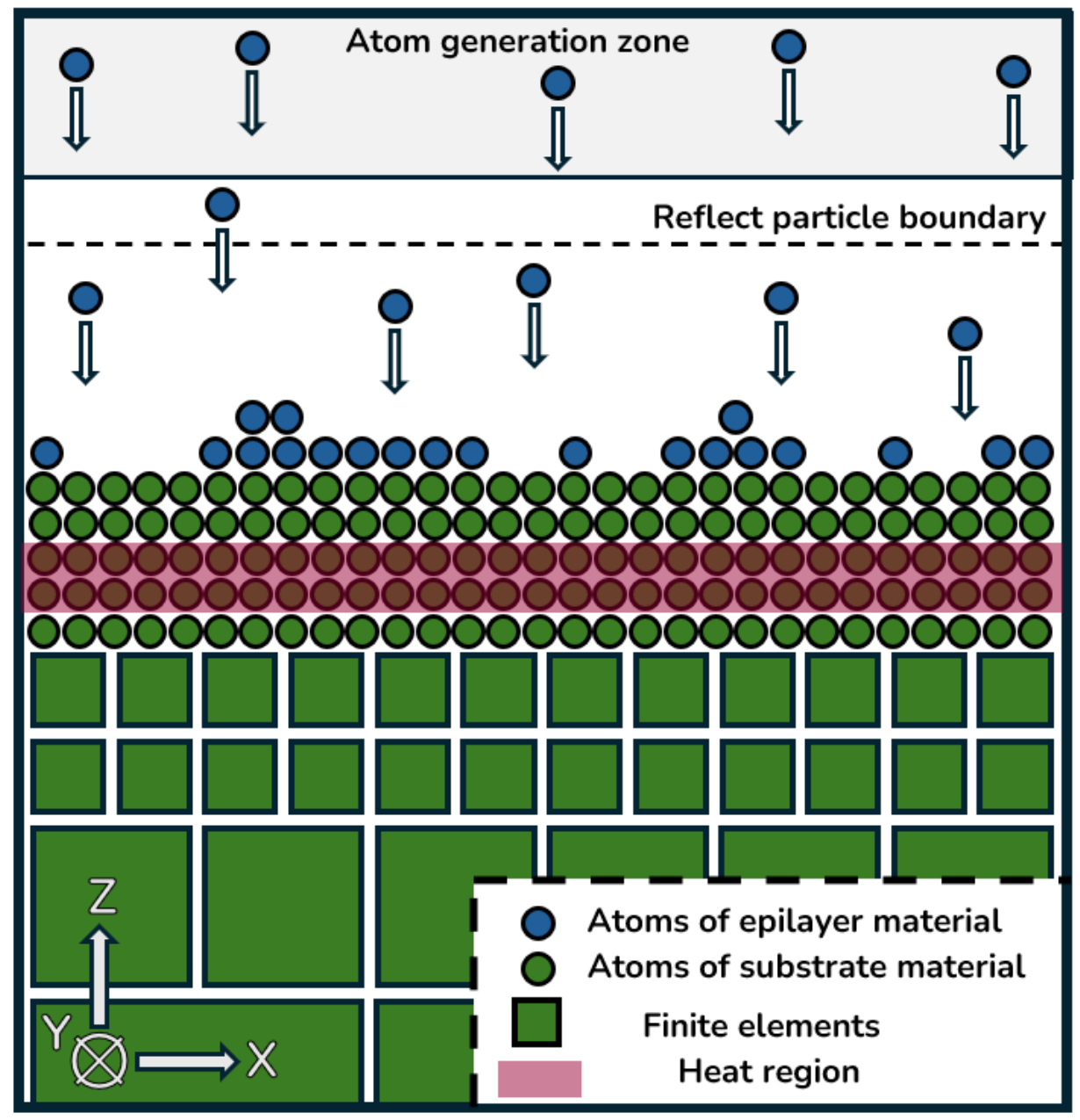

**Fig. 1. Schematic of CAC simulation of epitaxial growth**

For comparison purposes, coherent interfaces were created by simulations that constrain the substrate and overlayer in-plane lattice constants to the average of their equilibrium values. We note that such interfaces do not physically exist for lattice mismatched heterostructures such as the PbTe/PbSe system but are used here as a reference structure to isolate the effect of dislocations obtained through simulation of manufacturing processes. Both coherent and epitaxially grown interfaces were quenched to 0.7 K and energy minimized using the multi-stage annealing protocol consistent with that used for the directly bonded heterostructure.

The concept of surface or interfacial energy can be traced back to Gibbs, who described a "*dividing surface*" between "heterogeneous masses in contact" for which the "energy of the surface" is the difference between the total energy of a system and the sum of the energies of adjoining components separated by a dividing surface [26]. Strictly speaking, the difference between the total energy of a system and the sum of the energies of adjoining components separated by a dividing surface is the *interaction energy between components across the dividing surface*. Since an interface is a dividing surface and surface energy is defined as the work required to create a surface per unit area, we define the surface energy at an interface, i.e., the interfacial energy, as the work required to separate an interface into two free surfaces. In this work, we compute interfacial energies directly by measuring the interaction energy, $E_{int}$, across dividing surfaces defined as the total atomic interaction energy intersecting a surface per unit area. $E_{int}$ is evaluated by the sum of bond order contributions to the potential energy across the surface following the atomic-level flux formalism in our previous works, treating the problem in the form of line-plane intersection [27,28]. Physically, $E_{int}$ is a measure of the energy needed to break interatomic bonds across the surface at the interface and is equivalent to the work required to instantaneously separate the interface into two free surfaces.

We emphasize five key advantages of the computational method used to determine interfacial energies in this work. First, $E_{\mathrm{int}}$ is a physical quantity that applies generally to any dividing surface, including those in finite-sized heterostructures, interphase boundaries, cleavage planes, and interfaces containing misfit dislocations. Second, $E_{\mathrm{int}}$ can be calculated as a time or ensemble-averaged quantity or as an instantaneous quantity. Third, $E_{\mathrm{int}}$ is uniquely determined by the atomic structure of the interface. Fourth, the evaluation of $E_{int}$ requires no additional simulations or reference calculations. Fifth, $E_{\mathrm{int}}$ can be calculated during both equilibrium and non-equilibrium processes at finite temperature.

Verification of the interaction energy code was achieved by observing agreement between computed values of $E_{int}$ and surface energies obtained from slab-based methods. To quantify the surface energy at an epitaxial interface instantaneously, the epitaxial heterostructures containing the dislocation structures, as shown in Fig. 3, were held at growth temperature and $E_{int}$ was averaged across 100 ps to account for the effect of finite-temperature. The calculated energies were found to be time-independent with standard deviations $\leq 0.01$ eV nm$^{-2}$, confirming convergence at the given epilayer thickness.

Figure 2 presents the dislocation networks at the interfaces of PbSe/PbTe(100) and PbTe/PbSe(111) epitaxially grown heterostructures, as well as the atomic structures of the epilayer surfaces. We observe 3D dislocation structures with misfit-dislocations at the interface and threading dislocations extending towards the epilayer free surfaces. In addition, the growth mode of PbSe/PbTe(100) system is found to be layer-by-layer, i.e., Frank–van der Merwe 2D growth [29,30], while the growth mode of PbTe/PbSe(111) is the layer-plus-island, i.e., the Stranski–Krastanov, 2D+3D growth [31].

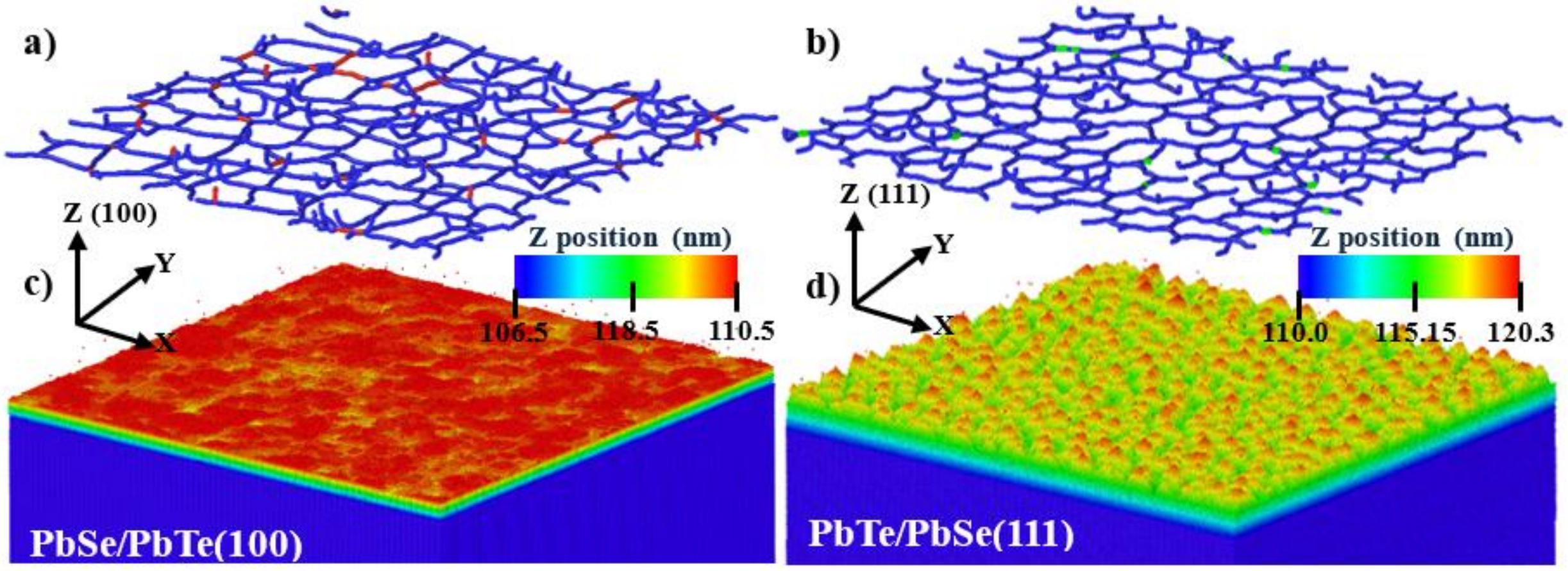

**Fig. 2. CAC simulation results of heteroepitaxy of PbSe/PbTe(100) and PbTe/PbSe(111) at 7.9 ML PbSe coverage and 7.9 ML PbTe coverage, respectively. a), b) dislocation networks and c) ,d) atomic displacements. The substrate sizes are ~ 100 × 100 × 100 nm$^3$. Visualized using OVITO Dislocation Extraction Algorithm (DXA)** [32,33]**.**

In Fig. 3, we compare dislocation structures obtained through simulations of direct bonding and with that of heteroepitaxy. As can be seen from Fig. 3, in (100) systems, directly bonded interfaces contain 2D regular square edge-dislocation networks, whereas epitaxial interfaces exhibit irregular 3D structures with mixed misfit and threading dislocations. Dislocation densities, obtained by normalizing total dislocation length by interfacial area, are found to be lower in epitaxial (100) systems (≈2.2–2.3 μm/μm$^2$) than that in the directly bonded interface (≈2.5 μm$^{-1}$). For PbTe-PbSe (111) systems, as shown in Fig. 3, the directly bonded interface forms a dense hexagonal network (≈2.3 μm$^{-1}$), and the epitaxial structures show contrasting densities between PbSe/PbTe (≈1.3 μm$^{-1}$) and PbTe/PbSe (≈2.1 μm$^{-1}$). The misfit dislocations are edge type with Burgers vectors along ⟨110⟩. Their slip planes are {100} for (100) interfaces and {111} for (111) interfaces, resulting in different characteristic shapes of the dislocation networks.

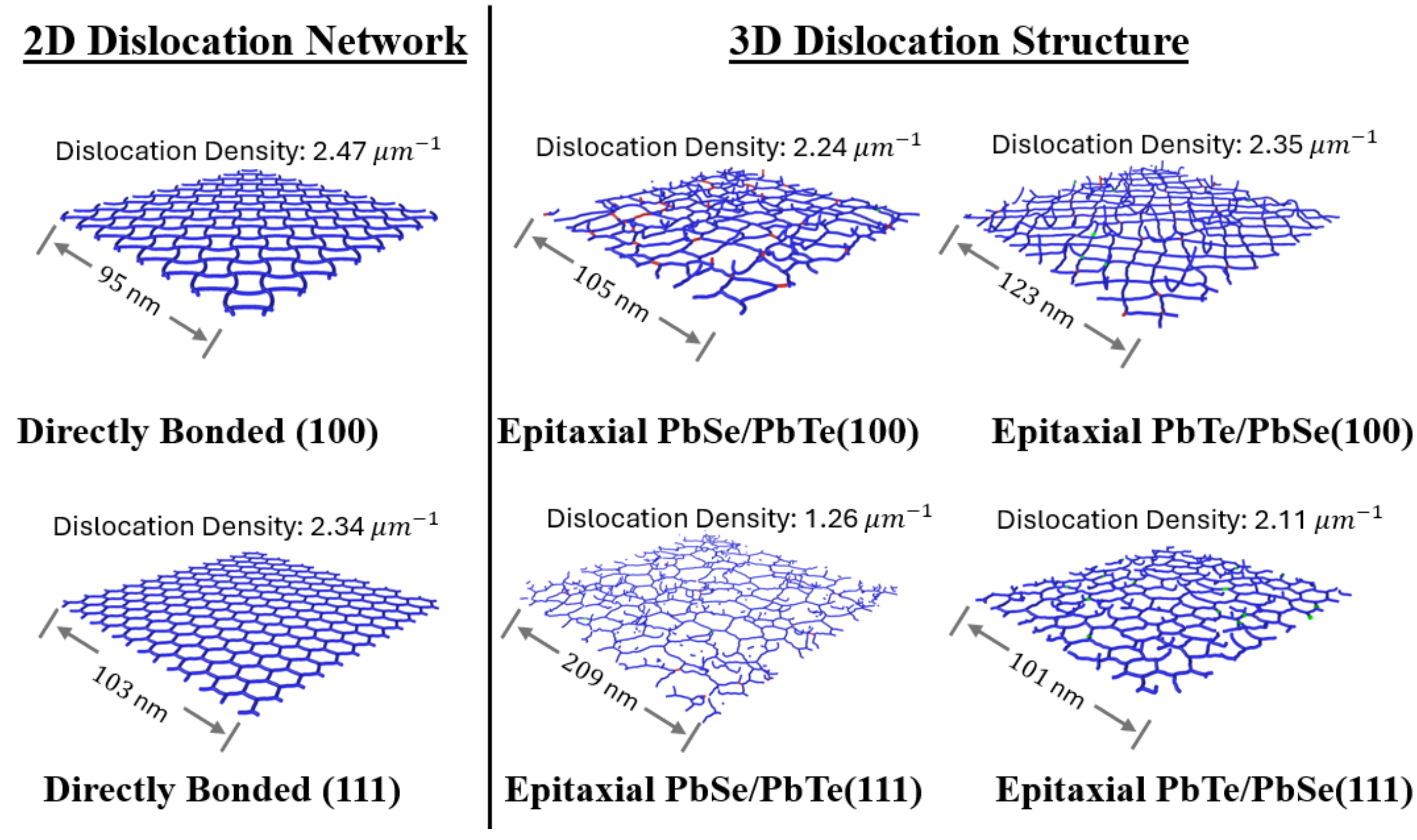

**Fig. 3. Dislocation networks of PbTe-PbSe interfaces obtained by CAC simulation of their manufacturing processes. The dislocation structures are visualized via OVITO (DXA)** [32,33].

**Table I** lists the computed surface energies of PbTe and PbSe single crystals at (100) and (111) interfaces. The (100) surfaces have ~3 times lower surface energy than the (111) surfaces, consistent with bond-counting arguments as (111) surface atoms lose three nearest neighbors whereas (100) surface atoms lose only one. The lower surface energy at the (100) surface may contribute to the formation of islands during epitaxy on (111) surfaces which are pyramids that expose the (100) facet, as observed in experiments [19,34,35]. We note that (111) surfaces are polar with Pb and Se/Te terminations, thus, the reported values correspond to the combined energy of each surface.

**Table I. Surface energies of PbTe and PbSe single crystals**

| Single-crystal interface | $E_{int}$ [eV/nm$^2$] |
| --- | --- |
| PbSe(100) | 3.69 |
| PbTe(100) | 3.33 |
| PbSe(111) | 13.25 |
| PbTe(111) | 10.59 |

**Table II** reports the interfacial energies, $E_{\text{int}}$, for various PbTe-PbSe(100) and PbTe–PbSe(111) interfaces. For the PbTe-PbSe(100) systems, the coherent interface has the highest surface energy, while the directly bonded semi-coherent interface with misfit dislocations has the lowest surface energy. The dislocation structure and density appear to correlate with interfacial energy. This is evidenced by two observations. First, epitaxially grown PbTe/PbSe(100) interfaces, which resemble the directly bonded interface in both dislocation structure and density, exhibit similarly low interfacial energies. Second, epitaxially grown PbSe/PbTe(100) interfaces, which have lower dislocation densities and more irregular dislocation patterning, retain higher interfacial energies.

Similar to interfaces in single crystals, the surface energies at PbTe-PbSe(111) interfaces are significantly larger than their (100) counterparts and show greater variation across interface types. For Pb-terminated interfaces, the interfacial energy of the directly bonded interface is close to that of the coherent interface, likely because the Pb-terminated interface is less chemically discontinuous than (100) interfaces, as both PbTe and PbSe share a common Pb layer. In contrast, the Se/Te-terminated directly bonded interface has substantially lower interfacial energy than the coherent interface. The annealed and instantaneous epitaxially grown (111) interfaces span a wide range of interfacial energies. A consistent trend is that the instantaneous epitaxial interfaces exhibit the lowest interfacial energies, suggesting that highly non-equilibrium conditions and the formation of complex three-dimensional dislocation networks may play a significant role in reducing the interfacial energy of (111) interfaces during epitaxial growth.

**Table II. Interfacial energies of various PbTe-PbSe Interfaces**

| Interface type | $E_{int}$ [eV/nm$^2$] | | |
|---|---|---|---|
| | (100) Interface | (111) Interface | |
| | | Pb-terminated | Se/Te-terminated |
| Coherent | 3.51 | 13.02 | 10.98 |
| Directly bonded | 2.85 | 12.69 | 9.48 |
| Annealed epitaxial PbTe/PbSe | 2.96 | 10.95 | 10.48 |
| Annealed epitaxial PbSe/PbTe | 3.37 | 9.96 | 10.27 |
| Instantaneous epitaxial PbTe/PbSe | 2.88 $\pm$ 0.003 | 10.57 $\pm$ 0.01 | 7.21 $\pm$ 0.008 |
| Instantaneous epitaxial PbSe/PbTe | 3.13 $\pm$ 0.002 | 9.19 $\pm$ 0.009 | 7.99 $\pm$ 0.004 |

**Table III** summarizes the percent difference in interfacial energy between coherent interfaces with directly bonded and epitaxially grown interfaces as $(E_{int}^{config} - E_{int}^{coh})/E_{int}^{config}$, where $E_{int}^{coh}$ is the interfacial energy of the coherent interface and $E_{int}^{config}$ is the interfacial energy of the interface obtained by the simulation of their respective manufacturing processes. For PbTe-PbSe(100) interfaces, the interfacial energy of the directly bonded interface is ~23% lower than the coherent interfaces with epitaxially grown interfaces exhibiting differences between -4.2% and -21.9%. For PbTe-PbSe Pb-terminated (111) interfaces, the interfacial energy of the directly bonded interface is only 2.5% lower with epitaxially grown interfaces exhibiting surface energies ~19 – 42% lower. In Se/Te-terminated (111) interfaces, the surface energies of directly bonded interfaces are ~16% lower than coherent interfaces, while epitaxially grown interfaces exhibit significant variation with ~ 5 – 52% lower surface energies than coherent interfaces.

**Table III. Percent difference in the interfacial energies of PbTe-PbSe coherent interfaces with directly bonded and epitaxially grown interfaces**

| Reference Configuration | Percent Difference [%] | | |
|---|---|---|---|
| | (100) Interface | Pb-terminated (111) Interface | Se/Te-terminated (111) Interface |
| Directly bonded | -23.2 | -2.5 | -15.8 |
| Annealed epitaxial PbTe/PbSe | -18.6 | -18.9 | -4.8 |
| Annealed epitaxial PbSe/PbTe | -4.2 | -30.7 | -6.9 |
| Instantaneous epitaxial PbTe/PbSe | -21.9 | -23.2 | -52.3 |
| Instantaneous epitaxial PbSe/PbTe | -12.1 | -41.7 | -37.4 |

Before concluding, we note that an alternative definition of interfacial energy exists in which it is defined as the excess free energy due to the presence of an interface. To distinguish this quantity from the definition of interfacial energy employed in this work, we refer to it as the excess interfacial free energy, $\gamma^*$. For a heterostructure consisting of two adjoining components, $A$ and $B$, separated by an interface of area $S$, the excess interfacial free energy is calculated as $\gamma^* = {}^{1}/_{S}\,[E_{A/B}^{tot} - E_{A}^{\mathrm{EQ}} - E_{B}^{\mathrm{EQ}}]$, where $E_{A/B}^{tot}$ is the total energy of the *A/B* heterostructure containing the interface, and $E_{A}^{\mathrm{EQ}}$ and $E_{B}^{\mathrm{EQ}}$ are the energies of the well-relaxed and equilibrated components $A$ and $B$ in isolation, i.e., in the absence of the interface [36,37]. **In Table IV**, we provide calculated excess interfacial free energies of various PbTe-PbSe heterostructures. We emphasize that, to be precise, excess interfacial free energy is a distinct concept and quantity from the work required to create surfaces, i.e., the surface or interfacial energy.

**Table IV. Excess interfacial free energies of various PbTe-PbSe heterostructures**

| Interface Type | $\gamma^*$ [eV/nm$^2$] | |
|---|---|---|
| | (100) Heterostructure | (111) Heterostructure |
| Coherent | 4.77 | 5.15 |
| Directly bonded | 1.51 | 1.47 |
| Annealed epitaxial PbTe/PbSe | 3.63 | 1.23 |
| Annealed epitaxial PbSe/PbTe | 3.97 | 1.66 |

As shown in **Table IV**, coherent heterostructures exhibit excess interfacial free energies more than three times larger than those of directly bonded interfaces, consistent with the substantial elastic strain energy stored in coherently constrained, lattice-mismatched systems. For (100) heterostructures, epitaxially grown interfaces exhibit $\gamma^*$ values intermediate between those of coherent and directly bonded interfaces, indicating partial strain relaxation via dislocation formation during growth. In contrast, for (111) heterostructures, epitaxially grown interfaces exhibit $\gamma^*$ values comparable to or lower than those of directly bonded interfaces. This suggests that three-dimensional island formation during the Stranski–Krastanov growth mode may play a significant role in reducing the excess interfacial free energy of epitaxially grown thin-films.

To recap, we have obtained the dislocation structures of various PbTe-PbSe interfaces through simulations of two different manufacturing processes: direct bonding and heteroepitaxy. Direct bonding produced interfaces with 2D misfit dislocation networks whereas heteroepitaxy produced interfaces with complex 3D dislocation structures. We determined the surface energies of these interfaces by computing the interaction energies across them. Relative to defect-free coherent interfaces, the directly bonded interfaces have up to 23% lower interfacial energy, whereas interfaces grown by heteroepitaxy can exhibit even larger reductions — up to 52%. This work demonstrated the significant effect of misfit and threading dislocations on interfacial energy.

The significance of this work may be further illustrated through a simple example. Consider Bauer's theory of growth mode [5], which was interpreted by van der Merwe in terms of interaction energies as [6,38]:

$$\Delta\gamma = E_{int}^{OO} - E_{int}^{\mathrm{OS}} \leq 0 \rightarrow \mathrm{FM}\ (layer\ by\ layer)$$
$$> 0 \rightarrow \mathrm{VM}\ (island\ formation)$$

where $\mathrm{E}_{int}^{\mathrm{OO}}$ is the "the work (per unit area of interface) needed to separate two half-crystals (of the growing crystal) from each other" and $\mathrm{E}_{\mathrm{int}}^{\mathrm{OS}}$ "the work (per unit area of interface) needed to separate a growing half-crystal from the semi-infinite substrate" [6]. If we consider the growth of PbTe on PbSe(111), using the coherent or directly bonded Pb-terminated interfaces to compute $E_{int}^{\mathrm{OS}}$, the theory predicts FM growth mode. In contrast, if we use the instantaneous epitaxial interface to compute $E_{int}^{\mathrm{OS}}$, the theory then predicts Volmer Weber (VM) [39] growth mode. Thus, the difference in the computed surface energies at an interface can lead to conflicting mechanistic or qualitative predictions on material behavior or phenomena.

*Acknowledgements:*

This work is based on research supported by the US National Science Foundation under Award Number DMR 2121895. The work for the computational code by Nicholas Taormina was supported by CMMI- 2349160, and the work of Yang Li was supported by CMMI-2054607. The computer simulations used Anvil and Expanse through allocation TG-DMR190008 from the Advanced Cyberinfrastructure Coordination Ecosystem: Services & Support (ACCESS) program.